\documentclass[aps,prd,superscriptaddress]{revtex4-2}
\usepackage[
unicode,
pdfauthor={Leron Borsten, Dimitri Kanakaris, 金炯錄},
pdftitle={Ersatz gravity and black-hole thermodynamics from Manin gauge theory with noncompact gauge group},
pdfsubject={hep-th,gr-qc,math-ph},
breaklinks=true,
final
]{hyperref}
\usepackage[T1]{fontenc}
\usepackage[unicode,final]{hyperref}
\usepackage{amsmath,amssymb,orcidlink,cleveref}
\usepackage[utf8]{inputenc}
\usepackage{CJKutf8}
\renewcommand\Im{\operatorname{im}}
\newcommand\intprod{\mathbin\lrcorner}
\newcommand\coloneqq{=}

\begin{document}
\title{Ersatz gravity and black-hole thermodynamics from Manin gauge theory with noncompact gauge group}

\author{Leron Borsten}
\email{l.borsten@herts.ac.uk}
\affiliation{Centre for Mathematics and Theoretical Physics Research, Department of Physics, Astronomy and Mathematics, University of Hertfordshire, Hatfield, Hertfordshire\ AL10 9AB, United Kingdom}
\author{Dimitri Kanakaris}
\email{d.kanakaris-decavel@herts.ac.uk}
\affiliation{Centre for Mathematics and Theoretical Physics Research, Department of Physics, Astronomy and Mathematics, University of Hertfordshire, Hatfield, Hertfordshire\ AL10 9AB, United Kingdom}
\author{Hyungrok Kim (\begin{CJK*}{UTF8}{bsmi}金炯錄\end{CJK*})}
\email{h.kim2@herts.ac.uk}
\affiliation{Centre for Mathematics and Theoretical Physics Research, Department of Physics, Astronomy and Mathematics, University of Hertfordshire, Hatfield, Hertfordshire\ AL10 9AB, United Kingdom}

\begin{abstract}
We show that a three-dimensional Manin gauge theory with gauge group \(\operatorname{SL}(2;\mathbb R)\) (i.e.\,Yang-Mills theory, the third-way theory, or the imaginary third-way theory) minimally coupled to Einstein gravity admits a dual interpretation as Einstein gravity with an exotic coupling to a Manin gauge theory, where the roles of dreibein/spin connection and field strength/gauge potential are interchanged.
The dual, or ersatz, gravitational metric \(\hat g_{\mu\nu}\sim\operatorname{tr}((\star F)_\mu (\star F)_\nu)\) is  a classical double copy of the gauge field strength \(F_{\mu\nu}\) (as opposed to the usual double copy of the gauge potential $A_\mu$).
If matter exclusively couples to  \(\hat g\) (for example, in a gravitational decoupling limit), then one can formulate black-hole thermodynamics with regard to the ersatz metric.  In particular, a black-hole solution for the ersatz metric \(\hat g_{\mu\nu}\) (made of Yang-Mills fields) radiates ersatz Hawking radiation and obeys the laws of black-hole thermodynamics.
\end{abstract}

\maketitle
\tableofcontents
\section{Introduction and summary}
\subsection{Background}
Gravity and gauge theories at first glance seem very different: gravity features spin-two excitations whereas gauge theories have spin-one excitations;
gravity couples to all matter universally, whereas gauge theories only couple to certain particles; gravity is perturbatively non-renormalisable, while for certain choices of global symmetries, gauge groups and matter fields, gauge theories are renormalisable or even super-renormalisable;  it seems that gravity is gravity, and gauge is gauge, and never the twain shall meet.
Yet, some recent (and not so recent) results challenge this view:
\begin{itemize}
\item Gravity features event horizons, which leads to it being inherently thermodynamic with phenomena such as Hawking radiation and black-hole thermodynamics. Although not strictly analogous,  Rindler horizons and Unruh radiation, for instance, are qualitatively similar and occur in quantum field theory sans gravity \cite{Fulling:1972md,Davies:1974th,Unruh:1976db}.
\item The above point is suggestive and, in fact, many non-gravitational systems are known that reproduce certain aspects of classical and quantum gravity, a phenomenon we call \emph{ersatz gravity}. 
This is in the spirit of  \emph{analogue} gravity (reviewed in \cite{Visser:2001fe,Barcelo:2005fc,Visser:2012rct}), but broader in the sense that we do not  make claims regarding  the feasibility of experimental realisations. 
In particular, Unruh's pioneering work \cite{Unruh:1980cg} proposed experimentally realisable analogue Hawking radiation in fluid systems. The pivotal observation was that Hawking radiation is not a property of general relativity specifically; rather, all that is typically required is a quantum field theory playing out on an effective geometry with a horizon.  By now there are numerous analogues  of Hawking radiation in non-relativistic acoustic systems \cite{Unruh:1994je,Visser:1998qn,Visser:2001fe} and other systems (see the surveys \cite{Barcelo:2005fc,Visser:2012rct}). For example, this effect has been reported to be experimentally observed in Bose-Einstein condensates \cite{Steinhauer:2015saa} (but see the discussion in \cite{Almeida:2022otk,Barcelo:2005fc}).

\item The double copy \cite{Bern:2008qj,Bern:2010ue,Bern:2010yg} (reviewed in \cite{Carrasco:2015iwa,Bern:2019prr,Borsten:2020bgv,Adamo:2022dcm,Bern:2022wqg}) relates scattering amplitudes of Yang-Mills theory to those of gravity; beyond scattering amplitudes, certain (but not all) solutions of Einstein gravity including some black holes can be constructed as double copies of solutions to Yang-Mills theory in a construction known as the classical double copy \cite{Monteiro:2014cda}.
Recently, it has been shown that Hawking radiation can be obtained as a double copy of certain processes in pure Yang-Mills theory \cite{Ilderton:2025aql,Aoude:2025jvt,Carrasco:2025bgu}. This raises the question of the extent to which Hawking radiation and, more generally, black-hole thermodynamics (see reviews \cite{Davies:1978zz,Wald:1999vt,Ross:2005sc,Wall:2018ydq})
are features specific to quantum gravity or whether there are ersatz versions of black-hole thermodynamics in the context of gauge theories.
\end{itemize}
In general, to what extent can gauge theory mimic or simulate aspects of gravity in a manner that is local (unlike AdS/CFT)?\footnote{Of course, under the AdS/CFT correspondence, large-\(N\) gauge theories show gravitational behaviour in a holographic sense including black holes and Hawking radiation. However, this correspondence is non-local in that the gravitational behaviour is on a different, higher-dimensional spacetime than that of the gauge theory itself.} The known analogue gravity systems typically  reproduce the \emph{kinematics} of gravity. That is, the fact that matter is coupled to an effective metric tensor describing a pseudo-Riemannian manifold. Hence, they reproduce only the kinematical parts of black-hole thermodynamics, such as Hawking radiation \cite{Unruh:1980cg,Visser:1998qn,Jacobson:2012ei,Visser:2012rct} and the second law of black-hole thermodynamics \cite{Hawking:1971tu} (requiring energy conditions; cf.~\cite{Ross:2005sc,Wall:2009wm}), which also hold for modified theories of gravity in addition to pure Einstein gravity. Indeed, the original derivation of Hawking radiation \cite{Hawking:1974rv} never uses Einstein's field equations. Consequently, the result also holds in modified metric theories of gravity as long as particles only couple to a metric field \(g_{\mu\nu}\) and even in non-gravitational systems where (quasi)particles experience an effective metric \(\hat g_{\mu\nu}\). If \(\hat g_{\mu\nu}\) has an effective horizon, it will  generate analogue Hawking radiation \cite{Unruh:1994je,Visser:1998qn,Visser:2001fe}.
On the other hand, the first law of black-hole thermodynamics does depend on the dynamics of gravity and changes for modified theories of gravity \cite{Ross:2005sc}. Here we put forward gauge-theory models of ersatz gravity that capture both the kinematics and dynamics, and  bring together elements of analogue gravity and the classical double copy. 

\subsection{Motivation}\label{ssec:motivation}
Let us first start with the kinematics of gravity. We seek to construct a gauge-invariant effective metric \(\hat g_{\mu\nu}\) out of gauge fields \(A^a_\mu\) to which matter, such as a scalar field \(\phi\), couples minimally:
\begin{equation}\label{eq:interaction-term}
    S=-\frac12\int\mathrm{d}^3x\sqrt{|\det\hat g|}(\hat g^{-1})^{\mu\nu}\partial_\mu\phi\partial_\nu\phi.
\end{equation}
The classical double copy suggest the obvious guess
\begin{equation}
    \hat g_{\mu\nu}\overset?\sim A^a_\mu A_{a\nu}.
\end{equation}
We leave the precise meaning of $\sim$ here ambiguous to accommodate the variety of possibilities.  For example, under the classical double copy the Coulomb potential generates the Schwarzschild solution \cite{Monteiro:2014cda}. Alternatively, for appropriate background spacetime metrics, one may introduce a bi-adjoint scalar and replace the pointwise product with a convolution. The gauge transformations then generate the correct residual diffeomorphisms  of   $\hat g_{\mu\nu}$ treated as a perturbation about the background metric \cite{Anastasiou:2014qba,Anastasiou:2018rdx,Borsten:2019prq,Borsten:2021zir}, avoiding this objection.

However, straightforwardly regarding $A^a_\mu A_{a\nu}$ as a map between off-shell fields   fails since it is not gauge-invariant. One may question whether the gauge-invariance of \(\hat g_{\mu\nu}\) is necessary as long as \eqref{eq:interaction-term} is gauge-invariant. This necessitates, however, that \(\phi\) must gauge-transform in a strange, non-canonical fashion. This scenario can be realised in three dimensions; Chern-Simons theories with noncompact gauge groups (namely, the isometry groups of Minkowski or (anti-)de~Sitter spaces) are known to be equivalent to Einstein gravity with a possible cosmological constant \cite{Achucarro:1986uwr,Witten:1988hc}, and in this case the corresponding ersatz metric \(\hat g_{\mu\nu}\) is not gauge-invariant. However, any matter coupled to the ersatz metric \(\hat g_{\mu\nu}\) must gauge-transform in a particular peculiar way involving derivatives, so that it is debatable whether they should be called `matter' at all; see \cref{sec:Chern-Simons}.

Instead, if one demands that \(\hat g_{\mu\nu}\) be gauge-invariant and built pointwise out of the gauge field alone, the simplest answer is
\begin{equation}
    \hat g_{\mu\nu}\sim F^a_{\mu\rho} F_{a \nu\sigma}g^{\rho\sigma},  
\end{equation}
where \(g^{\rho\sigma}\) is a background metric; more generally \(F\) can be replaced with any tensor that transforms as an adjoint field under gauge transformations rather than as a gauge connection. Note that, while this proposal still has a double copy form, it departs from the expectation that $F^a_{\mu\rho} F_{a\nu\sigma}+F^a_{\nu\sigma} F_{a\mu\rho}\sim R_{\mu\rho\nu\sigma}$ \cite{Anastasiou:2014qba}. 

The next task is to introduce ersatz gravity dynamics. That is,  \(\hat g_{\mu\nu}\) should evolve according to something resembling Einstein's field equations.
This task is harder but  possible in three dimensions at the cost of using a noncompact gauge group, where   Manin gauge theories \cite{Arvanitakis:2024dbu,Arvanitakis:2025nyy}, which include Yang-Mills theory, are known to be equivalent to Einstein gravity (with or without  a cosmological constant) coupled to a  background stress-energy tensor that breaks diffeomorphism invariance \cite{Borsten:2024pfz, Borsten:2024alh}.

The broken diffeomorphism symmetry can then be repaired by instead starting with a Manin theory coupled to Einstein gravity, which can be dualised 
into a \emph{bimetric} theory of gravity consisting of  the `true' metric \(g_{\mu\nu}\) and the ersatz metric \(\hat g_{\mu\nu}\) coming from the gauge fields of the Manin theory.\footnote{
    The end result is somewhat reminiscent of the bimetric massive theories of gravity \cite{Hassan:2011zd} (reviewed in \cite{Schmidt-May:2015vnx}) that naturally arise in the context of de~Rham-Gabadadze-Tolley massive gravity \cite{deRham:2010kj} that have attracted recent attention.
} (We refer to this as `duality', but note that it is unrelated to the AdS/CFT correspondence.) In this case,   gauge-invariant matter can be coupled to the ersatz metric rather than the true metric; then it is the event horizon of the ersatz metric, rather than the true metric, that radiates.

\paragraph{Summary of results.}
We propose a classical equivalence (at least perturbatively) 
between (1) Yang-Mills theory in three dimensions with gauge algebra  \(\mathrm T^*\mathfrak{sl}(2;\mathbb R)\cong \mathfrak{sl}(2;\mathbb R) \ltimes \mathfrak{sl}(2;\mathbb R)^*\)
coupled to Einstein gravity with zero cosmological constant and (2) Einstein gravity coupled to a  \(\mathrm T^*\mathfrak{sl}(2;\mathbb R)\) gauge field in an unusual fashion.
Under this duality, the components of the gluon \((A^a_\mu, B^{a}_\mu)\) are interchanged with the dreibein and the spin connection \((e^a_\mu,\omega^{a}_\mu)\). Which doublet  is `really' gravity, then, depends on whether uncharged matter couples to \((A^a_\mu, B^{a}_\mu)\) or to \((e^a_\mu,\omega^{a}_\mu)\). Matter will `experience' different spacetimes, depending on which field one chooses to couple to.

A consequence of the construction is that, indeed, an ersatz metric built out of gauge fields can radiate Hawking radiation in a double-copy-like fashion as
\begin{equation}\label{eq:ersatz-metric-ansatz}
    \hat g_{\mu\nu}\sim B^a_\mu B_{a\nu},
\end{equation}
where \(B^a_\mu\) is one of the components of the Manin-theory gauge field \(\mathbb A^{\mathbb a}_\mu=(A^a_\mu,B^a_\mu)\).
The ersatz metric is built as a double copy of the gauge field; if all matter couples to this ersatz metric rather than to the true metric, then indeed the gauge-theory ersatz black holes can radiate just like gravitational black holes.

\paragraph{Limitations.}
The Manin gauge theory used here necessarily uses a noncompact gauge group and hence is not an ordinary unitary gauge theory (one of the modes have a wrong-sign kinetic term),
unlike the Yang-Mills theory that appear in \cite{Ilderton:2025aql,Aoude:2025jvt,Carrasco:2025bgu}.

Furthermore, the coupling to the ersatz metric, from the perspective of the gauge theory, is non-minimal and may appear strange. (The coupling must be non-minimal because it is a coupling to gauge-invariant matter.)
In addition, the coupling is only well defined in the regime where the gauge field \(B^a_\mu\) is invertible; in particular, the coupling blows up if one tries to do perturbation theory close to the vacuum configuration with \(B^a_\mu=0\). (This vacuum configuration, however, is not natural from the dual perspective of bimetric gravity since it corresponds to the case where one of the two metrics is degenerate.)

Finally, this paper is restricted to a discussion of three-dimensional theories; a generalisation to higher spacetime dimensions is not immediate (but see \cite{Borsten:2024alh}).

\subsection{Organisation}
This paper is organised as follows. In \cref{sec:Chern-Simons}, we first review the gravitational reformulation of Chern-Simons theory with noncompact gauge groups in three dimensions, and explain why, if one tries to reproduce Hawking radiation, one can only couple to matter transforming in a peculiar way under gauge transformations. In \cref{sec:manin}, we review the Manin theory and its gravitational dual. Finally, in \cref{sec:bimetric}, we discuss Manin theory coupled to Einstein gravity and its dual as a bimetric theory of gravity and consequences for ersatz black-hole thermodynamics and ersatz Hawking radiation.

\paragraph{Conventions.}
We use the \(-++\) metric signature with the Levi-Civita symbol \(\varepsilon_{012}=-\varepsilon^{012}=1\). Symmetrisations \(\dotsm_{(\mu}\dotsm_{\nu)}\) and antisymmetrisations \(\dotsm_{[\mu}\dotsm_{\nu]}\) are always normalised. The notation \(\odot\) indicates a symmetrised product: if \(\alpha=\alpha_\mu\,\mathrm dx^\mu\) and \(\beta=\beta_\mu\,\mathrm dx^\mu\) are one-forms, then \((\alpha\odot\beta)_{\mu\nu}\coloneqq\alpha_{(\mu}\beta_{\nu)}\) is a symmetric tensor of rank two. We use differential-form notation freely; the operator \(\intprod\) denotes the interior product between a vector field and a differential form.

\section{Hawking radiation from Chern-Simons theory?}\label{sec:Chern-Simons}
In three spacetime dimensions, it is in fact possible to engineer not only the kinematics of general relativity but also the dynamics of general relativity using Chern-Simons theory \cite{Achucarro:1986uwr,Witten:1988hc}. In this case, one may hope to build the BTZ black hole from Chern-Simons theory and observe Hawking radiation.  Interestingly, three-dimensional Chern-Simon theory enjoys colour-kinematics duality \cite{Ben-Shahar:2021zww,Borsten:2022vtg,Borsten:2023ned}, and its double copy has been argued to be a theory of a `generalised metric perturbation' \cite{Bonezzi:2024dlv}, providing another, \emph{a priori} distinct, connection to gravity\footnote{In fact,  M2-brane effective world-volume theories are of Chern-Simons-matter type and are known to double-copy to supergravity \cite{Bargheer:2012gv,Huang:2012wr,Huang:2013kca}, suggesting that Chern-Simons theory itself double-copies to Einstein gravity \cite{Borsten:2022vtg,Borsten:2023ned}. This suggests that Chern-Simons theory can both be equivalent to gravity and double-copy to gravity!}.

\subsection{Noncompact gauge groups for Chern-Simons theory from Manin pairs}
An important class of Chern-Simons theories with noncompact gauge group are given by so-called Manin pairs associated to symmetric
spaces. Manin pairs consist of data \((\mathfrak{g}\hookrightarrow\mathfrak{d},\langle-,-\rangle)\), where \((\mathfrak{d},\langle-,-\rangle)\) is a Lie algebra \(\mathfrak{d}\) with a \(\mathfrak{d}\)-invariant, non-degenerate bilinear form \(\langle-,-\rangle\colon \mathfrak d\odot\mathfrak d\to\mathbb R\) of split signature, and where \(\mathfrak{g}\hookrightarrow\mathfrak{d}\) is a Lie subalgebra which is Lagrangian with respect to \(\langle-,-\rangle\), i.e.\ \(\langle-,-\rangle|_\mathfrak{g} = 0\) and \(\dim \mathfrak{d} = 2\dim\mathfrak{g}\). This notion originates in the theory of integrable systems (see the review \cite{10.1007/BFb0113695}).

Here, we further impose that \(\langle-,-\rangle\)  induces a symmetric pair decomposition \(\mathfrak{d} \cong \mathfrak{g} \oplus \mathfrak{g}^\ast\), 
\begin{subequations}
\begin{align}
	[\mathfrak{g},\mathfrak{g}] &\subseteq \mathfrak{g},
	&
	[t_a,t_b] &= f_{ab}{^c}t_c,
	\\
	[\mathfrak{g},\mathfrak{g}^\ast] &\subseteq \mathfrak{g}^\ast,
	&
	[t_a,\theta^b] &= -f_{ac}{^b}\theta^c,
	\label{eq:t theta}
	\\
	[\mathfrak{g}^\ast,\mathfrak{g}^\ast] &\subseteq \mathfrak{g},
	&
	[\theta^a,\theta^b] &= h^{abc}t_c,
\end{align}
\end{subequations}
where \((t_a)\) a \(\mathfrak{g}\)-basis, \((\theta^a)\) the dual \(\mathfrak{g}^\ast\)-basis, \((f_{ab}{^c})\) the \(\mathfrak{g}\)-structure constants and \((h^{abc})\) a totally antisymmetric, \(\mathfrak{g}\)-invariant tensor. \eqref{eq:t theta} is enforced by the \(\mathfrak{d}\)-invariance \(\langle[t,t],\theta\rangle = \langle t,[t,\theta]\rangle\), total antisymmetry of \((h^{abc})\) is a consequence the \(\mathfrak{d}\)-invariance \(\langle[\theta,\theta],\theta\rangle = \langle\theta,[\theta,\theta]\rangle\), and \(\mathfrak{g}\)-invariance of \((h^{abc})\) follows from the Jacobi identities involving \(t\theta\theta\) and \(\theta\theta\theta\).
Symmetric Manin pairs give rise to Chern-Simons theories
\begin{equation}
\label{eq:Manin CS action homogeneous case}
	S_\mathrm{CS}[\mathbb A] =S_\mathrm{CS}[A,B] = k\int B_a\wedge F^a + \frac{1}{6}h^{abc}B_a\wedge B_b\wedge B_c,
\end{equation}
where \(\mathbb A = A_\mu^a t_a\otimes\mathrm dx^\mu + B_{a\mu}\theta^a\otimes\mathrm dx^\mu\) the \(\mathfrak{d}\)-connection and \(F^a = \mathrm dA^a + \tfrac12[A,A]^a\) is the field strength for the \(\mathfrak{g}\)-connection \(A\).

Suppose we equip a Lie algebra \(\mathfrak{g}\) with an invariant non-degenerate bilinear form \(\kappa\colon \mathfrak{g}\odot\mathfrak{g}\to\mathbb R\), then we can always construct three natural examples of symmetric Manin pairs \((\mathfrak{g}\hookrightarrow\mathfrak{d},\langle-,-\rangle)\), 
\begin{itemize}
	\item the semi-direct sum \(\mathfrak{g}\hookrightarrow\mathfrak{d} = \mathfrak{g}^\ast \ltimes \mathfrak{g}\) where \(\mathfrak{g}^\ast\) is regarded as Abelian, \(\mathfrak{g}\) acts on \(\mathfrak{g}^\ast\) via the coadjoint representation,  \(\langle-,-\rangle\) is  the natural dual pairing, and \(h^{abc} = 0\),
	\item the diagonal embedding \(\mathfrak{g}\hookrightarrow\mathfrak{d} = \mathfrak{g}\oplus\mathfrak{g}\) given by \(t_a = t_a^L + t_a^R\) and \(\theta^a = \tfrac12 \kappa^{ab}(t^L_b - t^R_b)\), which gives rise to \(\langle-,-\rangle = \kappa^L - \kappa^R\) and \(h^{abc} = \frac14 f^{abc}\) (indices raised by \(\kappa\)), and
	\item the complexification \(\mathfrak{g}\hookrightarrow\mathfrak{d} = \mathfrak{g}_{\mathbb C} = \mathfrak{g} \oplus_{\mathbb R} \mathrm i\mathfrak{g}\) with \(t_a\) the real embedding, \(\theta^a = \frac12\kappa^{ab}\mathrm i t_b\), split signature product \(\langle-,-\rangle = 2\Im\kappa_{\mathbb C}\) and \(h^{abc} = -\frac14 f^{abc}\).
\end{itemize}
These three cases can be summarised as
\begin{align}
\label{eq:three cases}
	h^{abc} &= \frac{\nu}{4} f^{abc},
	&
	\nu &= 
	\begin{cases}
		+1 & \mathfrak{d} = \mathfrak{g} \oplus \mathfrak{g}
		\\
		\phantom{+}0 & \mathfrak{d} = \mathfrak{g}^\ast \ltimes \mathfrak{g}
		\\
		-1 & \mathfrak{d} = \mathfrak{g}_{\mathbb C}
	\end{cases}.
\end{align}
In particular, we note that the \(\nu = 0\) case reduces to non-Abelian \(BF\) theory.

\subsection{Chern-Simons/Einstein gravity duality}
\label{sec:Chern-Simons/Einstein gravity duality}
Let us review the well-known statement that Chern-Simons theory with certain noncompact gauge algebras is equivalent to three-dimensional Einstein gravity with a cosmological constant \cite{Achucarro:1986uwr,Witten:1988hc} in the language of Manin pairs.

We consider the three canonical Manin pairs built from the three-dimensional Lorentz algebra \(\mathfrak{g} \cong \mathfrak{so}(1,2) \cong \mathfrak{sl}(2;\mathbb R)\). Dualising  \([ab]\mapsto a\) (\(a,b=0,1,2\)),  the basis is given by \((t_{[ab]})\mapsto (t_a = -\frac12\varepsilon_{abc}t^{[bc]})\) with structure constants \(f_{ab}{^c} = \varepsilon_{ab}{^c} = \varepsilon_{abd}\eta^{dc}\), where \((\eta_{ab}) = \operatorname{diag}(-1,1,1)\) is the Minkowski  metric, and \(\varepsilon_{012} = -\varepsilon^{012} = 1\) the Levi-Civita symbol. Further taking \(\kappa_{ab} = \eta_{ab}\), we then find that
\begin{equation}\label{eq:iso(X)-manin-pairs}
	\mathfrak{d} = 
	\left\{
	\begin{aligned}
		&\mathfrak{sl}(2;\mathbb R)^{\oplus 2} 
		&
		&\cong \mathfrak{so}(2,2) 
		&
		&\cong \mathfrak{iso}(\operatorname{AdS}_{1,2}),
		& 
		\nu &= +1
		\\
		&\mathfrak{sl}(2;\mathbb R)^* \ltimes \mathfrak{sl}(2;\mathbb R) 
		&
		&\cong \mathrm T^*\mathfrak{sl}(2;\mathbb R)
		&
		&\cong \mathfrak{iso}(\mathbb R^{1,2}),
		& 
		\nu &= \phantom{+}0
		\\
		&\mathfrak{sl}(2;\mathbb C)
		& 
		&\cong \mathfrak{so}(1,3) 
		&
		&\cong \mathfrak{iso}(\operatorname{dS}_{1,2}),
		& 
		\nu &= -1
	\end{aligned},
	\right.
\end{equation}
where by \(\mathfrak{iso}(X)\) we denote the isometry algebra of the pseudo-Riemannian manifold \(X \in \{\operatorname{AdS}_{1,2},\mathbb R^{1,2},\operatorname{dS}_{1,2}\}\). Then the Manin pair inclusion \(\mathfrak{sl}(2;\mathbb R)\hookrightarrow\mathfrak{iso}(X)\) is the Lorentz subalgebra of the isometry algebra. This gives us Chern-Simons theories
\begin{equation}\label{eq:three cases for Lorentz}
	S_\mathrm{CS}[A,B] = k\int B_a\wedge F^a + \frac{\nu}{24}\varepsilon^{abc}B_a\wedge B_b\wedge B_c,
\end{equation}
where we use the same notation as in \eqref{eq:Manin CS action homogeneous case}.

Suppose that one rescales the gauge fields under the correspondence
\begin{equation}
\begin{aligned}
\label{eq:substitution}
    \hat M_\mathrm{Pl}\hat e^a &= kB^a,&
    \hat g_{\mu\nu} &= \eta_{ab}\hat  e^a_\mu\hat  e^b_\nu,&
    \hat \omega^{ab} &= -\varepsilon^{abc}A_c,&
    \hat \Lambda &= -\frac14k^{-2}\nu M_\mathrm{Pl}^2,
\end{aligned}
\end{equation}
where \(\hat M_{\mathrm{Pl}}\) is an arbitrary energy scale.
Then the action \eqref{eq:three cases for Lorentz} becomes
\begin{equation}
    S_{\mathrm{CS}}[\hat e,\hat\omega] = \frac12\hat M_\mathrm{Pl} \int \varepsilon_{abc}\Big(\hat e^a\wedge\hat  R^{bc}-\frac13\hat\Lambda\hat e^a\wedge\hat e^b\wedge\hat e^c\Big),
\end{equation}
which is a first-order formulation of three-dimensional Einstein gravity with a cosmological constant \(\hat \Lambda\), where
\begin{equation}
    \hat R^{ab} = \mathrm d\hat \omega^{ab} + \hat \omega^a{_c}\wedge\hat \omega^{cb}
\end{equation}
is the Riemann curvature tensor expressed as a two-form.
Under this correspondence, the dual metric is of the double-copied form
\begin{equation}
    \hat g_{\mu\nu}=(\hat e\odot\hat e)_{\mu\nu}=M_\mathrm{Pl}^{-2}k^{2}(B\odot B)_{\mu\nu}=M_\mathrm{Pl}^{-2}k^{2}B^a_\mu B_{a\nu}.
\end{equation}

\subsection{Matter couplings in Chern-Simons vs.\ gravity}
While pure Chern-Simons gauge theory and pure Einstein gravity with a cosmological constant are therefore equivalent (at least perturbatively\footnote{non-perturbatively, the theories may differ as to whether one sums over principal bundles or not, whether one should sum over orientations or spin structures, or whether one should allow degenerate metrics (non-invertible gauge fields); see e.g.\ the discussion in \cite{Borsten:2024pfz}.}), they diverge in how matter is to be `typically' coupled: a covariant coupling to gravity is very different from a covariant coupling to a gauge field. This has drastic consequences on the kinds of gauge transformations that matter should have. If matter transforms in a way suitable for coupling to gravity (i.e.\ it transforms in an expected manner under diffeomorphism), this corresponds to a bizarre transformation for gauge theory (i.e.\ a transformation \emph{not} labelled simply by a representation of the gauge group but instead involving derivatives). Conversely, if matter transforms in a way suitable for gauge theory (i.e.\ in a manner labelled by a representation of the gauge group), it is not suitable for coupling to gravity (i.e.\ does not transform as expected under diffeomorphism).

Explicitly, let us discuss what it takes for a gravitationally minimal coupling of a real scalar field to be gauge-invariant:
\begin{equation}\label{eq:Chern-Simons-matter}
    S = S_\mathrm{CS} -\int\sqrt{|\det\hat g|}\frac12(\hat g^{-1})^{\mu\nu}\partial_\mu\phi\partial_\nu\phi.
\end{equation}
For this to be gauge-invariant, \(\phi\) must transform under an infinitesimal diffeomorphism \(\xi=\xi^\mu\partial/\partial x^\mu\) as
\begin{equation}
    \delta\phi = \xi^\mu\partial_\mu\phi
\end{equation}
and be invariant under dreibein Lorentz transformations (since \(\phi\) does not carry any dreibein index \(^a\)). What does this correspond to in the language of gauge theory?

In Chern-Simons theory, an infinitesimal gauge parameter is a \(\mathfrak d\)-valued scalar field, which we may decompose as \(\mathbb c^{\mathbb a} = (\epsilon^a,\alpha^a{}_b)\). Let us define
\begin{equation}\epsilon^a=\xi\intprod \hat{e}^a.\end{equation}
Then under local Poincaré
\begin{equation}\delta \hat{e}^a=d\epsilon^a+\hat{\omega}^a{}_b\wedge\epsilon^b+\alpha^a{ }_b \hat{e}^b=\mathcal L_\xi \hat{e}^a-\xi\intprod T^a+\left(\alpha^a{ }_b-\xi\intprod\hat{\omega}^a{ }_b\right) \hat{e}^b,\end{equation}
where \(\alpha\) is the local Lorentz parameter and
\begin{equation}T^a = \mathrm d \hat{e}^a + \hat{\omega}^a{}_b\wedge \hat{e}^b\end{equation}
is the torsion. Since the interaction term does not explicitly depend on the spin connection \(\hat{\omega}\) but only on the dreibein \(\hat e\), the equations of motion enforce that \(T = 0\). Hence, on shell, diffeomorphisms parameterised by the vector field \(\xi\) correspond to infinitesimal gauge transformations where
\begin{align}
    \epsilon^a&=\xi\intprod \hat{e}^a,&
     \alpha^a{}_b&=\xi\intprod\hat{\omega}^a{ }_b,
\end{align}
where the latter shifts the local Lorentz transformation given by  \(\alpha^a{}_b\). Assuming the dreibein \(\hat e\) to be invertible, 
\begin{equation}
    \xi = \hat{e}^{-1}\epsilon.
\end{equation}
Thus, \(\phi\) must gauge-transform as
\begin{equation}\label{eq:scalar-diffeomorphism-transform}
    \delta \phi = \epsilon^a (\hat{e}^{-1})^\mu_a \partial_\mu\phi
\end{equation}
and be invariant under \(\alpha^a{}_b\) in order for \eqref{eq:Chern-Simons-matter} to be gauge-invariant on shell. Of course, this is just the expected general coordinate transformation rule for a scalar when using tangent indices.

However, from the original  gauge-theoretic point of view, \eqref{eq:scalar-diffeomorphism-transform} is very strange. Ordinarily matter should gauge-transform as
\begin{equation}
    \delta\phi^i = \mathbb c^{\mathbb a}R_{\mathbb a}{}^i{}_j\phi^j,
\end{equation}
where \(R_{\mathbb a}{}^i{}_j\) is a representation of \(\mathfrak d\) and \(i\) an index over the representation space. In particular, the gauge transformation is ultralocal, not merely local, in the sense that there are no derivatives of \(\phi\) involved. In contrast, \eqref{eq:scalar-diffeomorphism-transform} is not ultralocal. Rather, it is what you would expect of a local spacetime translation symmetry, as opposed to an internal symmetry.  Gauging a spacetime isometry group with matter minimally coupled as in \eqref{eq:Chern-Simons-matter} is not the same as a conventional gauge theory with matter, where in the latter case matter fields are sections of vector bundles associated to a principal $G$-bundle.

\subsection{Implications for Hawking radiation}
The Hawking radiation radiated by a black hole consists of gravitons and the quanta of other fields coupled to the metric. For the latter class, the usual derivations of Hawking radiation \cite{Hawking:1974rv,Hawking:1975vcx} assume that matter couples to the metric in the gravitationally minimal sense.

Suppose that one starts with pure Chern-Simons theory valued in \(\mathfrak g\) and constructs a solution dual to a BTZ black hole. Does it produce Hawking radiation? If one has pure Chern-Simons theory, the answer is, strictly speaking, no, since there are no gravitons or gluons that can be radiated; the theory lacks any degrees of freedom. Next, suppose that one has Chern-Simons theory coupled to some matter. Does the gauge-field configuration dual to a BTZ black hole radiate? The ordinary derivations of Hawking radiation go through only if the `matter' in question gauge-transforms sufficiently bizarrely; in particular, if the matter gauge-transforms in an ordinary (i.e.\ ultralocal) manner, then there is no reason to suppose that Hawking radiation would occur; a new computation, different from Hawking's, is needed to check this. In the subsequent sections we avoid this  observation altogether by providing generalisations that allow for the minimal coupling of matter to the ersatz metric in the usual manner required for Hawking radiation. 

\section{Review of three-dimensional Manin theories}\label{sec:manin}
We review  the construction in \cite{Borsten:2024pfz}, which relates three-dimensional Manin theories to three-dimensional Einstein gravity with a diffeomorphism breaking background stress-energy tensor, in view of the subsequent  extension to include a dynamical  metric in \cref{sec:bimetric}. In the latter case, the Manin theory is equivalent  to bimetric gravity.

A three-dimensional \emph{Manin theory} \cite{Arvanitakis:2024dbu,Arvanitakis:2025nyy} is a field theory associated to a Manin pair \((\mathfrak{g}\hookrightarrow\mathfrak d,\langle-,-\rangle)\) that is a deformation of a Chern-Simons theory valued in \(\mathfrak d\) with a non-derivative quadratic term:\footnote{This term appears similar to a Proca mass term but does not correspond to the masses of the propagating degrees of freedom, which are massless.}
\begin{equation}
    S[\mathbb A] = \int k\left\langle\mathbb A,\tfrac12\mathrm d\mathbb A + \tfrac16[\mathbb A, \mathbb A]\right\rangle
    -
    \tfrac12\left\langle\mathbb A,\star_g M\mathbb A\right\rangle
\end{equation}
where \(\mathbb A\) is a \(\mathfrak d\)-valued connection and \(M\colon \mathfrak d\to\mathfrak d\)
is a linear map of mass dimension \(1\) with kernel and image both equal to \(\mathfrak g\subset\mathfrak d\) such that
\begin{align}
    \langle My,z\rangle&=\langle y,Mz\rangle,&
    M[x,y]&=[x,My]
\end{align}
for \(x\in\mathfrak g\) and \(y,z\in\mathfrak d\). The Hodge star \(\star_g\) is taken with respect to a background (pseudo-)Riemannian metric \(g\).

We again consider the Manin pairs given in \eqref{eq:iso(X)-manin-pairs} and let $M\theta^a = \mu\eta^{ab}t_b$. Then the action of the Manin gauge theory is
\begin{equation}\label{eq:Manin_action}
\begin{aligned}
    S[A,B] &= \int k\operatorname{CS}[A,B]
    - \frac12\mu\eta^{ab}B_a\wedge\star_gB_b\\
    &=\int kB_a\wedge F^a + \frac1{24}\nu k\varepsilon^{abc}B_a\wedge B_b\wedge B_c 
    - \frac12\mu\eta^{ab}B_a\wedge\star_gB_b.
\end{aligned}
\end{equation}
where \(\operatorname{CS}[A,B]\coloneqq B_a\wedge F^a + \frac1{24}\nu\varepsilon^{abc}B_a\wedge B_b\wedge B_c\) is the Chern-Simons functional
or non-Abelian $BF$ functional when \(\nu = 0\), i.e.~\(X=\mathbb R^{1,2}\) as in \eqref{eq:three cases}.

When \(X=\mathbb R^{1,2}\),  i.e.\ \(\nu=0\), then the Manin theory is in fact three-dimensional Yang-Mills theory. When \(X=\operatorname{dS}_{1,2}\), the resulting theory is known as the third-way theory for \(\mathfrak{g} = \mathfrak{so}(1,2)\) \cite{Arvanitakis:2015oga} (reviewed in \cite{Deger:2021ojb}); when \(X=\operatorname{AdS}_{1,2}\), the resulting theory is known as the imaginary third-way theory for \(\mathfrak{g} = \mathfrak{so}(1,2)\) \cite{Arvanitakis:2024dbu,Arvanitakis:2025nyy}.

In the case \(X=\mathbb R^{1,2}\), \eqref{eq:Manin_action} is the first-order action of Yang-Mills theory \cite{Okubo:1979gt,McKeon:1994ds,Accardi:1997ps,Martellini:1997mu,Cattaneo:1997eh,Brandt:2015nxa,Frenkel:2017xvm,Brandt:2018avq,Brandt:2018wxe,Lavrov:2021pqh}, and by the equations of motion the auxiliary field \(B^a=(k\mu)\star_gF^a\) is the Hodge dual of the potential \(A\); in particular it transforms as the adjoint representation under the gauge algebra \(\mathfrak g\cong\mathfrak{sl}(2;\mathbb R)\). In the cases \(X\in\{\operatorname{AdS}_{1,2},\operatorname{dS}_{1,2}\}\) (i.e.\ \(\nu=\pm1\)), then the equations of motion imply that \(B^a\) is  given by
\begin{equation}
    B_a
    =
    \frac k\mu\eta_{ab}\star_g F^b + \frac{k^3}{8\mu^3}\nu\varepsilon_{abc}\star_g\big(\star_g F^b\wedge\star_g F^c \big)
    + \mathcal O\big((k/\mu)^4\nu^2F^3\big),
\end{equation}
which still transforms as the adjoint representation \cite{Arvanitakis:2015oga,Arvanitakis:2024dbu}. Therefore the ersatz metric \(\hat g_{\mu\nu}\sim \eta_{ab}B^a_\mu B^b_\nu\) is gauge-invariant in all cases.

The Manin theory obtained by adding the term \(MB\wedge\star B\) to Chern-Simons theory is equivalent to that of Einstein gravity with a cosmological constant and coupled to a background stress-energy tensor density \(T_{\mu\nu}\) under the correspondence \cite{Borsten:2024pfz}
\begin{equation}
    T^{\mu\nu} = k^{-2}\mu M_\mathrm{Pl}^2(g^{-1})^{\mu\nu}\sqrt{|\det g|}
\end{equation}
such that
\begin{equation}\label{eq:dual-Manin-action}
\begin{aligned}
	S[\hat e,\hat \omega] 
	=
	\frac12M_\mathrm{Pl} &\int \varepsilon_{abc}\Big(\hat e^a\wedge\hat  R^{bc}-\frac13\Lambda\hat  e^a\wedge\hat  e^b\wedge\hat  e^c\Big)
	\\
	{}-\frac12 &\int\mathrm d^3x\, T^{\mu\nu}\eta_{ab}\hat e^a_\mu\hat  e^b_\nu.
\end{aligned}
\end{equation}
Note that the background stress-energy tensor density \(T^{\mu\nu}\) breaks diffeomorphism symmetry.

\section{Manin theory coupled to gravity as bimetric gravity}\label{sec:bimetric}

In the previous section, we have considered a Manin theory with respect to a non-dynamical background metric \(g_{\mu\nu}\) such that the gravitational dual has broken diffeomorphism symmetry; it is a quantum field theory on fixed  curved background. To recover  diffeomorphism invariance, we now consider a Manin theory with fields \(A^a,B^a\) minimally coupled to Einstein gravity with fields \(e^a,\omega^a\), with Einstein-Hilbert and cosmological constant terms included. The action is therefore
\begin{equation}
\begin{aligned}
	S[A,B,e,\omega] 
	= \int
	&kB_a\wedge F^a + \frac1{24}\nu k\varepsilon^{abc}B_a\wedge B_b\wedge B_c - \frac12\mu\eta^{ab}B_a\wedge\star_gB_b
	\\
    &+ \frac12M_\mathrm{Pl} \varepsilon_{abc}\Big(e^a\wedge R^{bc}-\frac13\Lambda e^a\wedge e^b\wedge e^c\Big).
\end{aligned}
\end{equation}
This action simply describes the action of Manin theory minimally coupled to gravity.

\subsection{Dualisation into bimetric gravity}
Let us dualise the gauge field as before, using the substitution \cref{eq:substitution}. The action is then
\begin{equation}
\begin{aligned}
	S[e,\omega,\hat e,\hat \omega] = \int
	&\frac12M_\mathrm{Pl} \varepsilon_{abc}\Big(e^a\wedge R^{bc}-\frac13\Lambda e^a\wedge e^b\wedge e^c\Big)
	\\
	+{}&{}
	\frac12\hat M_\mathrm{Pl} \varepsilon_{abc}\Big(\hat e^a\wedge\hat R^{bc}-\frac13\hat\Lambda\hat e^a\wedge\hat e^b\wedge\hat e^c\Big)-\frac12 \lambda\hat g_{\mu\nu}g^{\mu\nu}\operatorname{vol}_g,
\end{aligned}
\end{equation}
where $R$ and $\hat R$ are the curvatures of the spin-connections $\omega$ and $\hat \omega$, respectively, and \(\operatorname{vol}_g\) is the volume form for \(g\), and
\begin{equation}
	\lambda = \frac{\mu \hat M_\mathrm{Pl}^2}{k^2}.
\end{equation}
Assuming that \(e\) and \(\hat e\) are both invertible, we may integrate out the spin connections \(\omega\) and \(\hat\omega\) to obtain an action only depending on the two metrics \(g\) and \(\hat g\):
\begin{equation}
\label{eq:action S(g,hat g)}
    S[g,\hat g] = \int\Big(\frac12M_\mathrm{Pl}R-\Lambda M_\mathrm{Pl}\Big)\operatorname{vol}_g
    + \Big(\frac12\hat M_\mathrm{Pl}\hat R-\hat\Lambda\hat M_\mathrm{Pl}\Big)\operatorname{vol}_{\hat g}
    -\frac12\lambda\hat g_{\mu\nu}g^{\mu\nu}\operatorname{vol}_g,
\end{equation}
where \(R\) and $\hat R$ now denote the scalar curvatures of \(g\) and $\hat g$, respectively.

The theory now has two independent Planck masses and cosmological constants, but the two metric fields do not enter the action in a fully symmetric fashion due to the coupling term \(\lambda\hat g_{\mu\nu}g^{\mu\nu}\operatorname{vol}_g\). Correspondingly, the equations of motion differ by a relative volume factor:
\begin{subequations}\label{eq:bimetric-eom}
\begin{align}
	\frac{2}{\sqrt{|\det g|}}\frac{\delta S}{\delta g^{\mu\nu}} &=
	M_{\mathrm{Pl}}(G_{\mu\nu}+\Lambda g_{\mu\nu}) - \lambda\left(\hat g_{\mu\nu}-\frac12\hat g_{\rho\sigma}g^{\rho\sigma}g_{\mu\nu}\right)
	= 0 \label{eq:bimetric-eom-1},
	\\
	\frac{2}{\sqrt{|\det \hat g|}}\frac{\delta S}{\delta\hat g_{\mu\nu}} &=
\hat M_{\mathrm{Pl}}(\hat G^{\mu\nu}+\hat \Lambda\hat g^{\mu\nu}) + \lambda g^{\mu\nu}\frac{\sqrt{|\det g|}}{\sqrt{|\det\hat g|}} = 0 \label{eq:bimetric-eom-2},
\end{align}
\end{subequations}
where it is understood that the Einstein tensor \(\hat G^{\mu\nu}\) is defined solely using \(\hat g\).

Now, either of the two equations in \eqref{eq:bimetric-eom} can be used to perturbatively solve for \(g\) in terms of \(\hat g\) (or vice versa), resulting in a theory of unimetric gravity with higher-derivative corrections.
Since we wish to examine the ersatz gravitational dynamics of \(\hat g\), we choose to integrate out \(g\) using \eqref{eq:bimetric-eom-1} (and retaining the \(\hat g\) ersatz Einstein equation \eqref{eq:bimetric-eom-2}), which implies
\begin{equation}
    \lambda\hat g_{\mu\nu}=M_{\mathrm{Pl}}(R_{\mu\nu}-2\Lambda g_{\mu\nu}).
\end{equation}
Then \(g\)  can be expressed in terms of \(\hat g\) as follows:
\begin{equation}\label{eq:g-in-terms-of-ĝ}
    g_{\mu\nu}=-(\lambda/2\Lambda M_{\mathrm{Pl}})\hat g_{\mu\nu}
    +(2\Lambda)^{-1}R_{\mu\nu}
    =
    -(\lambda/2\Lambda M_{\mathrm{Pl}})\hat g_{\mu\nu}
    +
    (2\Lambda)^{-1}\hat R_{\mu\nu}
    +
    \dotsb.
\end{equation}

Now \(g_{\mu\nu}\) can be integrated out in terms of \(\hat g_{\mu\nu}\), after which we thus obtain
\begin{equation}\label{eq:effective-action}
	S[\hat g]
	=
	\int\operatorname{vol}_{\hat g}\bigg[\frac12 \Big(\hat M_\mathrm{Pl} + \tilde\nu\Big|\frac{\lambda M_{\mathrm{Pl}}}{2\Lambda}\Big|^{\frac12}\Big)\hat R
	-
	\Big(\hat M\hat\Lambda + \tilde\nu\Big|\frac{\lambda^3}{2M_\mathrm{Pl}\Lambda}\Big|^\frac12\Big)\bigg] + \dotsb,
\end{equation}
where \(\tilde \nu = -\operatorname{sgn}(\Lambda)\), the higher-derivative terms have been omitted and the effective Planck mass \(\tilde M_\mathrm{Pl}\) and cosmological constant \(\tilde\Lambda\) are
\begin{align}
	\tilde{M}_\mathrm{Pl}
	&= 
	\hat M_\mathrm{Pl} + \tilde\nu\Big|\frac{\lambda M_{\mathrm{Pl}}}{2\Lambda}\Big|^{\frac12},\label{eq:effective-planck}\\
	\tilde{M}_\mathrm{Pl}\tilde{\Lambda}
	&= 
	\hat M_\mathrm{Pl}\hat\Lambda + \tilde\nu\Big|\frac{\lambda^3}{2M_\mathrm{Pl}\Lambda}\Big|^\frac12.\label{eq:effective-cosmological-constant}
\end{align}

Thus, in the regime where the higher-derivative terms can be neglected, we obtain Einstein gravity in three dimensions with respect to the ersatz metric \(\hat g\) (which ultimately derives from a double copy of the gluon field) with an effective Planck energy \(\tilde M_\mathrm{Pl}\) given by \eqref{eq:effective-planck} and an effective cosmological constant \(\tilde\Lambda\) given by \eqref{eq:effective-cosmological-constant} and with higher-order corrections.

\subsection{Other dualisations}
We briefly pause to note that   bimetric gravity is not the only possible dualisation: one can, for instance, dualise both the gauge field and the gravitational field to obtain a Chern-Simons field coupled non-minimally to gravity, or alternatively to only dualise the gravitational field to obtain two Chern-Simons theories coupled to each other.

That is, if one performs the substitution
\begin{align}
    \hat e^a &= kB^a/M_\mathrm{Pl},&
    \hat\omega^{ab} &= -\varepsilon^{abc}A_c,&
    \hat\Lambda &= -\frac14k^{-2}\nu M_\mathrm{Pl}^2,
    \\
    \hat B^a &= M_\mathrm{Pl}e^a/\hat k,&
    \hat A^a &= \frac12\varepsilon^{abc}\omega_{bc},&
    \hat\nu &= -\operatorname{sgn}\Lambda,\qquad
    \hat k =\begin{cases}
        \frac{M_\mathrm{Pl}}{2\sqrt{|\Lambda|}}&\text{if \(\Lambda\ne0\)}\\
        \text{arbitrary} & \text{if \(\Lambda=0\)},
    \end{cases}\notag
\end{align}
then the action is
\begin{equation}
\begin{aligned}
    S[\hat e,\hat\omega,\hat A,\hat{B}] = \int
    &\hat k\hat B_a\wedge\hat F^a + \frac1{24}\hat\nu\hat k\varepsilon^{abc}\hat B_a\wedge\hat B_b\wedge\hat B_c
    \\
    &+ \frac12M_\mathrm{Pl} \varepsilon_{abc}\Big(\hat e^a\wedge\hat R^{bc}-\frac13\hat\Lambda\hat e^a\wedge\hat e^b\wedge\hat e^c\Big)
    \\
    &-\frac12\mu\eta^{ab}M_{\mathrm{Pl}}k^{-2}\hat k\hat g_{\mu\nu}(\hat B^{-1})^\mu_c (\hat B^{-1})^\nu_d\eta^{cd}\operatorname{vol}_{\hat B\odot\hat B}.
\end{aligned}
\end{equation}
This action now describes a Chern-Simons theory coupled to Einstein gravity with cosmological constant \(\hat\Lambda\) in a non-minimal, non-polynomial coupling.

Similarly, if one only dualises only the gravitational field, then one obtains two copies of Chern-Simons theory coupled via an interaction term
\begin{equation}
    \int -\frac12\hat kM_{\mathrm{Pl}}^{-1}\mu \eta^{ab}B_{a\mu}B_{b\nu}(\hat B^{-1})^\mu_c (\hat B^{-1})^\nu_d\eta^{cd}|\det B|\,\mathrm d^3x.
\end{equation}
\subsection{Matter coupling and ersatz black-hole thermodynamics}
In three dimensions, a no-black-hole theorem \cite{Ida:2000jh,Mao:2023xxb} states that the existence of black holes requires a negative cosmological constant assuming that matter satisfies the dominant energy condition. In the present case, at least if we assume that \(M_{\mathrm{Pl}}\) is sufficiently small so that the higher-order terms are negligible and matter satisfies the dominant energy conditions, we may assume that perturbations of generalisations of the BTZ black hole with matter survive. In this case, does the horizon of the ersatz black hole built out of gauge fields radiate? We argue that, if matter couples to \(\hat g\) appropriately, then the answer is yes.
 
The characteristic feature of gravity is that it couples to every field universally (including itself): every field has a kinetic term, which is written with the metric. Suppose, now, that matter couples to the ersatz metric \(\hat g\) rather than to the true metric \(g\) directly; for example, for a scalar field,
\begin{equation}
\begin{split}
    S 
    &= -\frac12\int \mathrm d^3x\,\sqrt{|\det(B\odot B)|} ~ ((B\odot B)^{-1})^{\mu\nu}\partial_\mu\phi\partial_\nu\phi
    \\
    &\propto
    -\frac12\int \mathrm d^3x\,\sqrt{|\det\hat g|} ~ (\hat g^{-1})^{\mu\nu}\partial_\mu\phi\partial_\nu\phi.
\end{split}
\end{equation}
Note that \(\phi\) is gauge-invariant; in other words, fields without exotic gauge transformations can straightforwardly couple to the ersatz metric, unlike the case of pure Chern-Simons theory in \cref{sec:Chern-Simons}.
If all matter couples to \(\hat g\) rather than to \(g\) (i.e.~in a gravitational decoupling limit), in a field configuration with horizons for \(g\) and for \(\hat g\), it is the horizons for \(\hat g\) that radiate Hawking radiation, rather than those of \(g\) (if any).

In particular, in the limit where the higher-derivative terms in \eqref{eq:effective-action} are small, we obtain Einstein gravity with some higher-derivative corrections (that would be generated by the renormalisation group flow in any case), so that black-hole solutions for \(\hat g\) such as the BTZ black hole exist. Even if there is also a horizon for \(g\), in general the horizons for \(g\) and \(\hat g\) would be located at different places and with different specific gravities; the usual argument for Hawking radiation implies that the thermodynamic properties of Hawking radiation will follow those of \(\hat g\) rather than those of \(g\).

Similarly, the derivation of the second law of black-hole thermodynamics \cite{Hawking:1971tu}  follows, as long as the matter coupling obeys appropriate energy conditions and the higher-order corrections remain small enough to not violate them (cf.\ \cite{Ross:2005sc,Wall:2009wm} for the appropriate energy conditions needed).
 
Furthermore, since the dynamics of \(\hat g\) are those of Einstein gravity with higher-order corrections, we reproduce not only the kinematics of Einstein gravity but also the approximate dynamics of Einstein gravity. That is, the zeroth and first laws of thermodynamics hold in approximate form for in the limit where \(M_{\mathrm{Pl}}\) is small \footnote{
It has been recently shown that the third law of black-hole thermodynamics \cite{Bardeen:1973gs,Israel:1986gqz}, which states that extremal black holes cannot be realised from a non-extremal configuration, admits smooth counterexamples for Einstein gravity coupled to reasonable kinds of matter \cite{Kehle:2022uvc,Kehle:2023eni,Kehle:2024vyt,Marin:2024rxt,Gadioux:2025unn}.}.

\section{Conclusions}

Having seen how to construct ersatz metrics that reproduce the main features of black hole thermodynamics in three spacetime dimensions, it is natural to ask about higher dimensional generalisations \footnote{We thank a referee for prompting us to address this in more detail.}, especially in four spacetime dimensions. 

In 4d there is a very natural,  analogous, AKSZ gauge theory realisation of gravity given in \cite{Borsten:2024alh}.  It is very close to the 3d case, i.e.~there is an ASKZ topological gauge theory deformed by a Manin pair ``mass'' term that is classically equivalent to 4d general relativity. However, in this case the ersatz metric is  given by the classical double copy of $A$ and not $B$, which in 4d is related to the curvature $B\sim R$. Thus, in 4d we na\"ively run into the problems outlined in \autoref{ssec:motivation}. Moreover, in this case the Manin pair term required to relate the topological AKSZ model to \emph{dynamical} 4d general relativity  is  (somewhat surprisingly)  itself topological \footnote{This works because part of the gauge algebra is explicitly broken.}, so there is no additional  coupling to a second metric. Thus the double copy metric can be interpreted as an ersatz metric in the absence of gravity, or as the true metric in the absence of matter, but we cannot have both at once. Thus, this picture somewhat departs from our 3d analysis (although, it is not necessarily uninteresting for that reason). Finally, we note in \cite{Borsten:2024alh} that there is an alternative Manin deformation that does introduce a background metric, which could be promoted to a dynamical field yielding some kind of bimetric gravity theory. However, the two gravitational sectors are non-minimally coupled in this case and the resulting action is harder to interpret. For example, there is a would-be Palatini action for Einstein–Hilbert gravity, except that the Ricci ``scalar'' involves both metrics.

Finally, although we have been cautious to make no claims that such double copy realisations of black hole thermodynamics have laboratory realizations, we end by noting that there are  bimetric low-energy effective theory descriptions of fractional quantum Hall states that capture  the gapped Girvin-Macdonald-Platzman mode \cite{PhysRevX.7.041032}. This presents a possible route to analogue gravity scenarios for the theory presented here.

\section*{Acknowledgements}
Leron Borsten and Hyungrok Kim thank Christopher D. White\textsuperscript{\orcidlink{0000-0002-6752-2394}} for helpful discussion. We are grateful to Giandomenico Palumbo for bringing the bimetric descriptions of  fractional quantum Hall states to our attention. 
Leron Borsten is grateful for the hospitality of the Theoretical Physics group, Blackett Laboratory, Imperial College London.

\newcommand\cyrillic[1]{\fontfamily{Domitian-TOsF}\selectfont \foreignlanguage{russian}{#1}}

\bibliography{biblio}
\end{document}